\documentclass{article}
\usepackage{spconf,amsmath,graphicx,hyperref}
\usepackage{bm}
\usepackage{amsfonts}
\usepackage{multirow,makecell}
\usepackage{algorithmic,algorithm}
\usepackage{enumitem}

\title{ParaGSE: Parallel Generative Speech Enhancement with Group-Vector-Quantization-based Neural Speech Codec}
\name{Fei Liu, Yang Ai$^*$ \thanks{$^*$Corresponding author. This work was funded by the National Nature Science Foundation of China under Grant 62301521.}}
\address{National Engineering Research Center of Speech and Language Information Processing,\\ 
University of Science and Technology of China, Hefei, China\\
{\small \tt \ fliu215@mail.ustc.edu.cn, yangai@ustc.edu.cn}}
\begin{document}
\ninept
\maketitle
\begin{abstract}
Recently, generative speech enhancement has garnered considerable interest; however, existing approaches are hindered by excessive complexity, limited efficiency, and suboptimal speech quality. 
To overcome these challenges, this paper proposes a novel parallel generative speech enhancement (ParaGSE) framework that leverages a group vector quantization (GVQ)-based neural speech codec.
The GVQ-based codec adopts separate VQs to produce mutually independent tokens, enabling efficient parallel token prediction in ParaGSE. 
Specifically, ParaGSE leverages the GVQ-based codec to encode degraded speech into distinct tokens, predicts the corresponding clean tokens through parallel branches conditioned on degraded spectral features, and ultimately reconstructs clean speech via the codec decoder. 
Experimental results demonstrate that ParaGSE consistently produces superior enhanced speech compared to both discriminative and generative baselines, under a wide range of distortions including noise, reverberation, band-limiting, and their mixtures. 
Furthermore, empowered by parallel computation in token prediction, ParaGSE attains about a 1.5-fold improvement in generation efficiency on CPU compared with serial generative speech enhancement approaches.

\end{abstract}
\section{Introduction}

In real-world scenarios, speech is frequently impaired by distortions such as noise, reverberation, and bandwidth limitations \cite{benesty2008microphone}. 
Speech enhancement (SE) addresses these challenges by refining degraded speech, minimizing interference, and restoring clarity, ultimately improving both the listening experience and performance in applications like communication \cite{rabiner1995impact,mahdi2009advances} and speech recognition \cite{wang2019bridging,chen2015integration}. 

With the rise of deep learning, neural network–based SE has become the mainstream approach, surpassing traditional methods in enhancing speech quality. 
Early neural SE models, e.g., DEMUCS \cite{defossez2020real}, directly minimized the difference between predicted and raw clean speech using regression losses, but their performance was unsatisfactory. 
Later, discriminative SE methods were introduced, incorporating generative adversarial networks (GANs) to improve speech quality. 
Time-domain discriminative SE methods, e.g., SEGAN \cite{pascual2017segan}, use fully convolutional networks to predict clean speech from noisy input with adversarial loss, but they still suffer from limited performance and high computational cost. 
To address these challenges, frequency-domain discriminative SE methods have gained popularity. 
For example, CMGAN \cite{abdulatif2024cmgan} reconstructs clean speech waveforms by predicting complex spectra, while MP-SENet \cite{lu2023mp,lu2025explicit} enhances performance by explicitly predicting amplitude and phase spectra through parallel modeling. 

\begin{figure*}[t]
    \centering
    \includegraphics[width=0.99\linewidth]{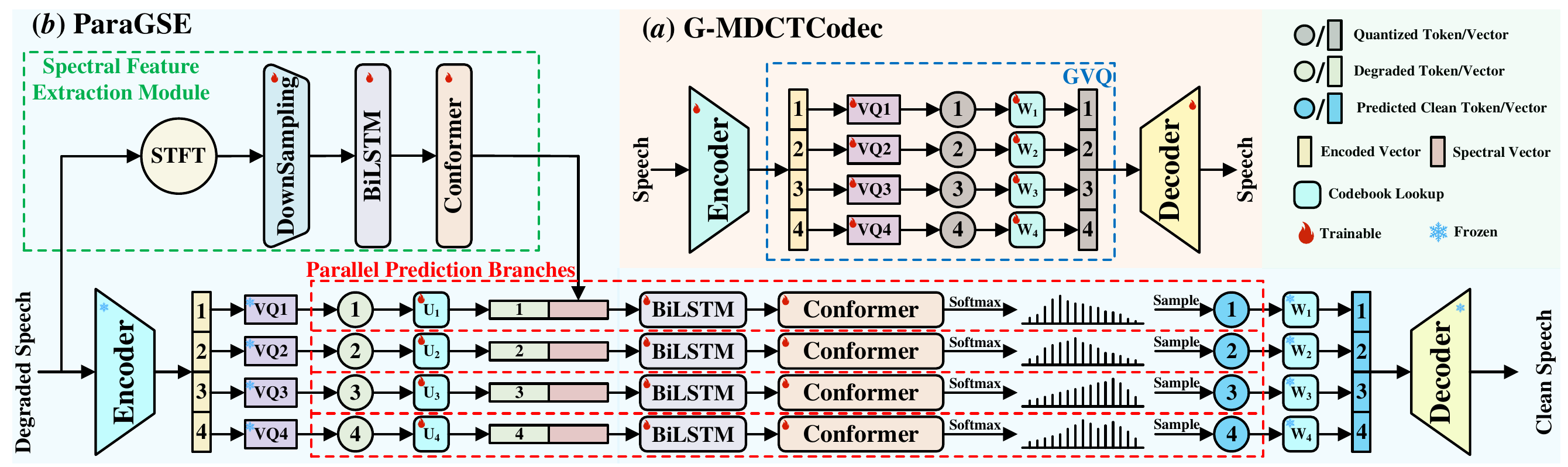}
    \caption{Details of (a) G-MDCTCodec and (b) ParaGSE, with $N=4$ as an example.}
    \label{fig:train}
\end{figure*}

Recently, rapid advancements in neural speech coding have driven the development of generative SE methods, aiming to leverage discrete tokens quantized by neural speech codecs to enhance speech quality \cite{wang2024speechx,xue2024low,wang2024selm,yang2024genhancer,yao2025gense}. 
In generative SE methods, the typical approach involves using a neural network to predict discrete tokens of clean speech from degraded input, followed by the reconstruction of clean speech through a codec decoder. 
The optimization objective is shifted to a token-level classification loss, rather than a regression loss or adversarial loss.
However, this class of methods remains in its emergent phase, confronting various challenges related to efficiency, complexity, and performance.
For example, GenSE \cite{yao2025gense} uses a large language model (LLM) to perform clean token prediction for decoding clean speech, which undoubtedly leads to high model complexity and low generation efficiency. 
Genhancer \cite{yang2024genhancer}, though not relying on LLMs, leverages a residual vector quantizer (RVQ)-based codec, resulting in autoregressive clean token prediction that, in turn, constrains computational efficiency. 
Additionally, both GenSE and Genhancer require semantic token inputs from pre-trained models, which also adds to their complexity. 

Therefore, this paper proposes ParaGSE, a novel parallel generative SE method supported by a group vector quantizer (GVQ)-based neural speech codec, which operates without the need for LLMs or semantic token assistance, enabling efficient parallel computations. 
This work presents two main contributions. 
First, we introduce G-MDCTCodec, a novel GVQ-based codec derived from our previously proposed MDCTCodec \cite{jiang2024mdctcodec}, which supports the parallel token prediction and training of ParaGSE through grouping and independent quantization. 
Another contribution is the development of a novel parallel token prediction framework in ParaGSE, serving the generative SE domain. 
In this framework, the G-MDCTCodec' encoder extracts independent tokens from the degraded speech, which are then processed by lightweight parallel prediction branches to output clean tokens. 
Finally, the clean speech is reconstructed through the G-MDCTCodec's decoder. 
Both objective and subjective experiments demonstrate that our proposed ParaGSE can restore higher-quality speech across various distortion types (e.g., noise, reverberation, band-limiting, and their mixtures), outperforming both generative and discriminative SE baselines. 
Additionally, thanks to parallel computation, ParaGSE delivers a notable efficiency advantage, running over 1.5$\times$ faster than serial generative SE on CPU.


\section{Proposed Method}

The proposed ParaGSE leverages G-MDCTCodec to enable parallel token prediction for generative SE. 
The structures of both components are depicted in Figure 1, with a detailed description provided below.


\vspace{-1.2mm}
\subsection{G-MDCTCodec: GVQ-based Neural Speech Codec}
\vspace{-1mm}
\label{sec: G-MDCTCodec}

As shown in Figure \ref{fig:train} (a), G-MDCTCodec primarily consists of an encoder, a GVQ, and a decoder. 
G-MDCTCodec determines the performance upper bound of ParaGSE, with all its components integrated and contributing to the training targets. 
A detailed description of the components of G-MDCTCodec is provided below. 
\vspace{-1mm}
\begin{itemize}[leftmargin=*]
\item {\textbf{Encoder \& Decoder:}} We retain the encoder and decoder architecture of MDCTCodec \cite{jiang2024mdctcodec}, using 1D ConvNeXt v2 blocks \cite{woo2023convnext} as the backbone to encode and decode the modified discrete cosine transform (MDCT) spectrum of the input speech. 
\item {\textbf{GVQ:}} The GVQ groups the encoded features along the dimension axis and quantizes each group using an independent VQ. 
Compared to the RVQ used in most codecs \cite{jiang2024mdctcodec,defossezhigh,kumar2023high,ai2024apcodec}, GVQ features a parallel rather than serial structure. 
Additionally, in contrast to the serial-parallel hybrid structure of grouped RVQ (GRVQ) used by HiFi-Codec \cite{yang2023hifi}, GVQ is more efficient and better suited for generative SE.
Assume one-frame encoded vector output by the encoder is $\bm{e} \in \mathbb{R}^K$, it is first divided into $N$ groups along the dimension axis, resulting in $N$ feature vectors $\bm{e}_n \in \mathbb{R}^{K/N}$, $n = 1, \dots, N$, where $K$ denotes vector dimension. 
Each grouped vector $\bm{e}_n$ is quantized using its corresponding VQ, where the codevector from the associated codebook $\mathbb{W}_n = \{\bm{w}_n^{(m)} \in \mathbb{R}^{K/N} \mid m = 1, \dots, M\}$ with the smallest Euclidean distance to $\bm{e}_n$ is selected as the quantized vector, and its index is regraded as the token, i.e.,
\begin{equation}
    \hat{\bm{e}}_n,d_n=\arg\min_{\bm{w}_n^{(m)},m} \| \bm{e}_n - \bm{w}_n^{(m)} \|_2,
\end{equation}
where $\hat{\bm{e}}_n\in \mathbb{R}^{K/N}$ and $d_n\in\{1,\dots,M\}$ respectively denote the quantized vector and token, and $M$ is the codebook size.
Finally, the quantized vectors from all groups are concatenated along the dimension axis to produce the final quantized result $\bm{\hat{e}}\in\mathbb R^K$, i.e.,
\begin{equation}
\label{equ: concat}
\bm{\hat{e}}=\left[\hat{\bm{e}}_1^\top,\hat{\bm{e}}_2^\top,\dots,\hat{\bm{e}}_N^\top \right]^\top,
\end{equation}
which is then decoded into speech. 

\item {\textbf{Training Criteria:}} During the training of G-MDCTCodec, we retain the adversarial loss, MDCT spectrum loss, and mel-spectrogram loss from MDCTCodec \cite{jiang2024mdctcodec}. 
An additional GVQ loss is included to optimize the learnable codebooks and reduce quantization error, defined as the sum of the mean squared errors (MSE) between the input and output of each VQ, i.e.,
\begin{equation}
    \mathcal{L}_{\mathrm{GVQ}} 
= \sum_{n=1}^N \mathbb{E}_{(\hat{\bm{e}}_n,\,\bm{e}_n)} 
\left\| \hat{\bm{e}}_n - \bm{e}_n \right\|_{F}^{2},
\end{equation}
where $\left\| \cdot \right\|_{F}$ is the Frobenius norm.
\end{itemize}

\subsection{ParaGSE: Parallel Generative Speech Enhancement}

As shown in Figure \ref{fig:train} (b), ParaGSE consists of a spectral feature extraction module $\phi^{(SFEM)}$, $N$ parallel prediction branches $\phi^{(PPB)}_1,\dots,\phi^{(PPB)}_N$, as well as the components of G-MDCTCodec. 
ParaGSE first feeds the degraded speech into the G-MDCTCodec's encoder and GVQ to obtain degraded tokens. 
Conditioned on spectral features from the spectral feature extraction module, the parallel prediction branches transform these degraded tokens into clean ones in parallel, which are then decoded by the G-MDCTCodec's decoder to reconstruct the clean speech.
The details of each component are as follows.
\begin{itemize}[leftmargin=*]
\item {\textbf{Spectral Feature Extraction Module:}} Taking one-frame feature as an example, the spectral feature extraction module $\phi^{(SFEM)}$ derives spectral feature vector $\bm{\hat{s}}\in \mathbb{R}^{C}$ from the degraded speech segment $\bm{y}\in \mathbb{R}^T$, i.e.,
\begin{equation}
    \bm{\hat{s}}=\phi^{(SFEM)}(\bm{y}),
\end{equation}
where $C$ denotes spectral feature dimension. 
$T$ denotes the speech segment length, which also corresponds to the downsampling factor of this module and the G-MDCTCodec's encoder.
Specifically, $\bm{y}$ is first transformed by a short-time Fourier transform (STFT) to obtain the amplitude and phase spectra, where the frame shift of STFT is $w_s$.
These two spectra are concatenated and passed through a downsampling module consisting of multiple convolutional layers with stride $>1$, which reduce the time resolution by a factor of $R$. 
Thus, we have $T = w_s\cdot R$.
After downsampling, two bidirectional long short-term memory (BiLSTM) layers are applied to capture global dependencies, followed by a Conformer block \cite{gulati2020conformer} that further models both global and local information, yielding the spectral feature $\bm{\hat{s}}$.
\item {\textbf{Parallel Prediction Branches:}} Taking the $n$-th branch $\phi^{(PPB)}_n$ as an example, it takes as input the token $d_n^{(y)}$ generated from $\bm{y}$ by the $n$-th VQ of the G-MDCTCodec's GVQ, together with $\bm{\hat{s}}$ as an auxiliary input, and outputs the clean token $\hat{d}_n^{(x)}$, i.e.,
\begin{equation}
    \hat{d}_n^{(x)}=\phi^{(PPB)}_n(d_n^{(y)};\hat{\bm{s}}).
\end{equation}
Specifically, the degraded token $d^{(y)}_n$ is first mapped to a latent vector $\bm{v}_n \in \mathbb{R}^C$ by looking up a trainable embedding table (i.e., codebook) $\mathbb{U}_n = \{\bm{u}_n^{(m)} \in \mathbb{R}^C \mid m = 1, \dots, M\}$.
After concatenating the latent vector $\bm{v}_n$ with the spectral feature vector $\hat{\bm{s}}$ to form $\hat{\bm{o}}_n\in\mathbb R^{2C}$, it is processed by two BiLSTM layers and one Conformer block, producing the output vector $\hat{\bm{z}}_n\in\mathbb R^M$.
Then, $\hat{\bm{z}}_n$ is passed through a softmax layer to compute the probability distribution $\hat{\bm{p}}_n^{(x)} \in \mathbb{R}^{M}$, i.e.,
\begin{equation}
    \hat{\bm{p}}_n^{(x)}=softmax(\hat{\bm{z}}_n).
\end{equation}
The predicted clean token $\hat{d}_n^{(x)}$ is obtained via maximum-value sampling (i.e., argmax) from the distribution $\hat{\bm{p}}_n^{(x)}=[\hat{p}_{n,1}^{(x)},\dots,$ $\hat{p}_{n,m}^{(x)},\dots,\hat{p}_{n,M}^{(x)}]^\top$, i.e.,
\begin{equation}
    \hat{d}_n^{(x)}=arg\max_m(\hat{p}_{n,m}^{(x)}).
\end{equation}
Finally, the predicted clean token $\hat{d}_n^{(x)}$ is restored to vector $\hat{\bm{e}}_n^{(x)}$ through the codebook $\mathbb{W}_n$ of the $n$-th VQ in G-MDCTCodec's GVQ. 
The vectors from all $N$ branches are then concatenated according to Eq. \ref{equ: concat} and fed into the decoder to reconstruct the clean speech.
\item {\textbf{Training Criteria:}} In ParaGSE, all components of G-MDCTCodec are frozen, and training is performed only on the spectral feature extraction module and $N$ parallel prediction branches. 
Specifically, the target clean speech $\bm{x} \in \mathbb{R}^T$ is first encoded and quantized by G-MDCTCodec to produce the target clean tokens $d_1^{(x)},\dots,d_N^{(x)}$. 
These tokens are individually converted into one-hot target probability distributions $\bm{p}_1^{(x)},\dots,\bm{p}_N^{(x)}$, which are then used to compute the cross-entropy loss with the predicted distributions $\hat{\bm{p}}_1^{(x)},\dots,\hat{\bm{p}}_N^{(x)}$, i.e.,
\begin{equation}
    \mathcal L=\sum_{n=1}^{N} CrossEntropy(\hat{\bm{p}}_n^{(x)},\bm{p}_n^{(x)}).
\end{equation}
\end{itemize}

\section{Experiments}
\subsection{Experimental Settings}

In our experiments\footnote{Codes and speech samples are available at: \url{https://anonymity225.github.io/ParaGSE/}.}, we constructed a clean speech dataset from the VoiceBank corpus \cite{valentini2016investigating}, comprising 23,075 utterances from 56 speakers for training and 824 utterances from 2 unseen speakers for testing, with a sampling rate of 16 kHz. 
We constructed task-specific degraded datasets from the clean speech corpus, incorporating three types of distortions: noise, reverberation, and mixed distortions (noise + reverberation + band-limiting). 
\begin{itemize}[leftmargin=*]
\item {\textbf{Denoising:}} For the denoising task, the noisy dataset was constructed by adding noise from the DEMAND dataset \cite{thiemann2013diverse} to the clean speech. 
For training, 10 types of noise were used with signal-to-noise ratios (SNRs) ranging from 0 to 15 dB in 5 dB steps, while for testing, 5 unseen noise types were applied with SNRs ranging from 2.5 to 17.5 dB in 5 dB steps.

\item {\textbf{Dereverberation:}} For the dereverberation task, we constructed the reverberant dataset using the room impulse response (RIR) dataset from the DNS Challenge \cite{dubey2024icassp}, which contains 248 real RIRs and about 60,000 simulated RIRs. 
Reverberant speech was generated by randomly selecting RIRs and convolving them with clean speech, ensuring that the RIRs used in the test set were unseen during training.

\item {\textbf{Mixed Distortion Suppression:}} For the mixed distortion suppression task, reverberation was first added to the clean speech, followed by noise, with both steps following the aforementioned procedures.
Finally, the speech signals were downsampled to 8 kHz (i.e., bandwidth reduction) to construct the mixed-distortion dataset.

\end{itemize}


The G-MDCTCodec in ParaGSE employed 4 VQs (i.e., $N=4$), each with a codebook size of 256 (i.e., $M=256$) and codevector dimension of 8 (i.e., $K=32$). 
All other configurations were kept the same as in MDCTCodec \cite{jiang2024mdctcodec}.
The spectral feature extraction module in ParaGSE adopted an STFT with a frame length of 320, a frame shift of 40 (i.e., $w_s=40$), and an FFT size of 1024. 
The subsequent downsampling module consisted of 3 convolutional layers, each with a kernel size of 3 and stride of 2, and had an overall downsampling factor of 8 (i.e., $R=8$).
Both the spectral feature extraction module and the parallel prediction branches employed Conformer blocks with 512 channels (i.e., $C=512$) and 8 attention heads. 
We trained the ParaGSE using an AdamW optimizer on a single Nvidia A800 GPU, with $\beta_{1}=0.9,\beta_{2}=0.95,$ and a weight decay of 0.01 for 100 epochs. 
The initial learning rate was set to 0.0005, with a cosine annealing strategy for decay and a warm-up training scheduler for the first 10k steps.

We compared ParaGSE against several representative baselines, including the early time-domain regression SE model DEMUCS \cite{defossez2020real}, the frequency-domain discriminative SE models CMGAN \cite{abdulatif2024cmgan} and MP-SENet \cite{lu2025explicit}, as well as the generative SE model Genhancer \cite{yang2024genhancer}. 
For fair comparison, DEMUCS, CMGAN, and MP-SENet were reproduced using their official implementations, while Genhancer was reimplemented based on the descriptions in its paper, as no official code is available.

\vspace{-1mm}
\subsection{Feasibility Study}
\vspace{-1mm}
\begin{table}[t]
    \centering
    \caption{Comparison of decoded speech quality between MDCTCodec and G-MDCTCodec.}
    \begin{tabular}{cccc}
    \hline

    \hline

         Codec & \textbf{LSD$\downarrow$} & \textbf{STOI$\uparrow$} & \textbf{VISQOL$\uparrow$}  \\ \hline
         \textbf{MDCTCodec} & 0.76 & \textbf{0.94} & \textbf{4.42} \\
         \textbf{G-MDCTCodec} & \textbf{0.75} & \textbf{0.94} & 4.39 \\ 
         \hline

         \hline

    \end{tabular}
    \label{tab:code}
\end{table}

As described in Section \ref{sec: G-MDCTCodec}, G-MDCTCodec determines the upper bound of the enhanced speech quality in ParaGSE. 
Therefore, the feasibility of ParaGSE relies on G-MDCTCodec achieving sufficiently high coding quality. 
In this section, we compared the coding results of G-MDCTCodec with its predecessor, the RVQ-based MDCTCodec \cite{jiang2024mdctcodec}.
We adopted commonly used evaluation metrics for speech coding, including log spectral distance (LSD), short-time objective intelligibility (STOI), and the virtual speech quality objective listener (ViSQOL). 
The experimental results are presented in Table \ref{tab:code}. 
We can observe that the decoded speech quality of G-MDCTCodec is comparable to that of MDCTCodec, indicating that although the VQs in GVQ are independent with weaker correlations compared to RVQ, this does not lead to a loss in speech quality. This serves as the prerequisite for enabling parallel token prediction in ParaGSE and verifies the feasibility of parallel generative SE.


\subsection{Comparisons with Baseline SE Methods}

\begin{table}[t]
    \centering
    \caption{Objective experimental results of ParaGSE versus baseline SE methods across three SE tasks.}
    \resizebox{\columnwidth}{!}{
    \begin{tabular}{cccccc}
    \hline

    \hline
         SE Task & Model & \textbf{LSD$\downarrow$} & \textbf{NISQA$\uparrow$} & \textbf{DNSMOS$\uparrow$} & \textbf{UTMOS$\uparrow$} \\ \hline
         \multirow{5}*{Denoising} & \textbf{DEMUCS} & 1.17 & 3.57 & 3.07 & 3.59 \\
         & \textbf{CMGAN} & \textbf{0.74} & \textbf{4.67} & 3.23 & \textbf{4.01} \\
         & \textbf{MP-SENet} & 0.79 & 4.66 & 3.23 & 4.00 \\
         & \textbf{Genhancer} & 1.10 & 3.40 & 2.65 & 2.73 \\ 
         & \textbf{ParaGSE} & 0.95 & 4.57 & \textbf{3.26} & 3.95 \\ \hline
         \multirow{5}*{Dereverberation} & \textbf{DEMUCS} & 1.78 & 1.67 & 2.68 & 1.39 \\
         & \textbf{CMGAN} & \textbf{0.72} & 4.45 & 3.13 & 3.54 \\
         & \textbf{MP-SENet} & 0.91 & 4.33 & 3.23 & 3.41 \\
         & \textbf{Genhancer} & 1.09 & 2.87 & 1.90 & 1.70 \\
         & \textbf{ParaGSE} & 0.84 & \textbf{4.60} & \textbf{3.26} & \textbf{3.86} \\ \hline
         & \textbf{DEMUCS} & 2.27 & 0.96 & 1.72 & 1.44 \\
         Mixed & \textbf{CMGAN} & 1.41 & 3.34 & 3.00 & 2.28 \\
         Distortion & \textbf{MP-SENet} & 1.62 & 3.40 & 3.10 & 2.27 \\
         Suppression & \textbf{Genhancer} & 1.40 & 3.47 & 2.01 & 1.95 \\
         & \textbf{ParaGSE} & \textbf{1.10} & \textbf{4.64} & \textbf{3.30} & \textbf{3.80} \\ 
         \hline

         \hline
    \end{tabular}}
    \label{tab:se}
\end{table}

To compare ParaGSE with baseline SE methods, we adopted objective and subjective metrics for evaluation. 
For the objective part, we employed LSD and three non-intrusive metrics widely used in SE, namely NISQA \cite{mittag2021nisqa}, DNSMOS \cite{reddy2022dnsmos}, and UTMOS \cite{saeki2022utmos}, to assess overall speech quality. 
For the subjective part, we conducted ABX preference tests on Amazon Mechanical Turk to compare ParaGSE against the two strongest baselines, CMGAN and MP-SENet. 
In each test, 20 utterances enhanced by the compared models were randomly selected from the test set and evaluated by at least 30 native English-speaking listeners. 
Listeners judged which sample in each pair had better quality or indicated no preference (N/P). We reported the average preference score and the $p$-value from a $t$-test to assess significance.

The objective and subjective experimental results are shown in Tables \ref{tab:se} and \ref{tab:subject}, respectively. 
For both the denoising and dereverberation tasks, ParaGSE consistently outperformed the time-domain model DEMUCS and the generative SE model Genhancer across all objective metrics, as shown in Table \ref{tab:se}.
Compared with the discriminative models CMGAN and MP-SENet, ParaGSE achieved comparable performance on the three non-intrusive objective metrics in the denoising task and even outperformed them in the dereverberation task. 
However, on the intrusive LSD metric, ParaGSE lagged behind CMGAN and MP-SENet, which can be attributed to the fact that generative models aim to approximate the true distribution rather than directly reconstruct clean speech. 
For further evidence, Table \ref{tab:subject} shows the subjective results. 
In the denoising task, ParaGSE performed on par with CMGAN ($p=0.625$) and significantly better than MP-SENet ($p<0.01$), while in the dereverberation task it clearly outperformed both baselines ($p<0.01$).
These results indicate that, despite weaker performance on intrusive metrics like LSD, generative models can better capture perceptually important characteristics. 
It demonstrates that ParaGSE delivers competitive performance in the classical tasks of denoising and dereverberation. 



For the mixed distortion suppression task, ParaGSE’s advantage became more pronounced, with both objective and subjective results in Tables \ref{tab:se} and \ref{tab:subject} showing significant improvements over all baselines ($p<0.01$). 
Notably, it surpassed even the strong baselines CMGAN and MP-SENet across multiple metrics including LSD, while listeners consistently preferred the perceptual quality of ParaGSE-enhanced speech.
These findings demonstrate its strong adaptability and robustness in handling complex distortions, underscoring its potential for practical applications.

\subsection{Efficiency Analysis Experimental Results}
\vspace{-1mm}

\begin{table}[t]
    \centering
    \caption{Subjective ABX preference scores (\%) of ParaGSE versus CMGAN and MP-SENet across three SE tasks.}
    \resizebox{\columnwidth}{!}{
    \begin{tabular}{cccccc}
    \hline

    \hline
         SE Task & \textbf{ParaGSE} & \textbf{CMGAN} & \textbf{MP-SENet} & N/P & $p$ \\ \hline
         \multirow{2}*{Denoising}& 32.63 & 34.08 & - & 33.29 & 0.625 \\
         & \textbf{66.84} & - & 17.76 & 15.40 & \textbf{$\bm{<}$ 0.01}  \\ \hline
         \multirow{2}*{Dereverberation}& \textbf{42.41} & 32.04 & - & 25.55 & \textbf{$\bm{<}$ 0.01} \\
         & \textbf{56.22} & - & 24.46 & 19.32 & \textbf{$\bm{<}$ 0.01}  \\ \hline
         Mixed Distortion & \textbf{77.63} & 15.00 & - & 7.37 & \textbf{$\bm{<}$ 0.01} \\
         Suppression & \textbf{82.97} & - & 10.00 & 7.03 & \textbf{$\bm{<}$ 0.01}  \\ 
         \hline

         \hline
    \end{tabular}}
    \label{tab:subject}
\end{table}

\begin{table}[t]
    \centering
    \caption{RTF analysis results of ParaGSE, Genhancer, and SerialGSE for mixed distortion suppression task. Here ``$a\times$" denotes $a\times$ real time.}
    \resizebox{\columnwidth}{!}{
    \begin{tabular}{ccccc}
    \hline

    \hline
         Model & Type & \textbf{UTMOS$\uparrow$} & \textbf{RTF (GPU)$\downarrow$} & \textbf{RTF (CPU)$\downarrow$} \\ \hline
         \textbf{Genhancer} & Serial & 1.95 & 0.1160 (8.6$\times$) & 0.6535 (1.5$\times$) \\
         \textbf{SerialGSE} & Serial & 3.21 & 0.0308 (32.5$\times$) & 0.0696 (14.4$\times$) \\
         \textbf{ParaGSE} & Parallel & \textbf{3.80} & \textbf{0.0228 (43.9$\times$)} & \textbf{0.0466 (21.5$\times$)} \\ 
         \hline

         \hline
    \end{tabular}}
    \label{tab:eff}
\end{table}

To verify the efficiency advantage of ParaGSE’s parallel prediction strategy, we compared it with serial-prediction-based generative SE methods, e.g., Genhancer. 
However, Genhancer adopts DAC \cite{kumar2023high} and relies on additional semantic token inputs, rendering the comparison with ParaGSE inequitable. 
To address this, we constructed a serial baseline, SerialGSE, derived from ParaGSE. 
Compared with ParaGSE, SerialGSE differs only in replacing G-MDCTCodec with MDCTCodec \cite{jiang2024mdctcodec} and adopting an RVQ-based sequential prediction scheme. 
Specifically, the clean token of each VQ is predicted from its noisy token and the clean tokens of all preceding VQs, while each prediction branch retains the same structure as in ParaGSE.

We used real-time factor (RTF) to measure generation efficiency, defined as the time required to generate one second of speech, and evaluated it on Nvidia A800 GPU and Intel(R) Xeon(R) Silver 4314 CPU. 
The experimental results for the mixed distortion suppression task are presented in Table \ref{tab:eff}. 
Genhancer shows the lowest efficiency, attributed to its serial prediction scheme, the high complexity of DAC, and reliance on semantic token inputs. 
Its quality (measured by UTMOS) is also suboptimal, likely because it fails to leverage RVQ dependencies and was originally designed for denoising \cite{yang2024genhancer}, making it less effective in handling complex distortions. 
Compared with SerialGSE, ParaGSE has a clear efficiency advantage, being 1.5 times faster on CPU due to parallel prediction and acceleration. 
It also achieves higher speech quality as measured by UTMOS. 
These results show that ParaGSE’s parallel framework improves both generation efficiency and perceptual quality, underscoring its practicality for real-time and resource-constrained applications.

\vspace{-1mm}
\section{Conclusion}
\vspace{-1mm}

This paper introduces ParaGSE, a novel parallel generative SE model built on a GVQ-based neural speech codec. 
The codec employs separate VQs to generate independent tokens, enabling ParaGSE to encode degraded speech, predict clean tokens in parallel conditioned on spectral features, and reconstruct enhanced speech through the decoder. 
Extensive experiments demonstrate that ParaGSE consistently surpasses both discriminative and generative baselines across diverse distortions, while also achieving about a 1.5-fold efficiency gain on CPU over serial approaches. 
In future work, we plan to further improve ParaGSE and investigate its effectiveness in broader application scenarios, including more challenging real-world conditions.


\vfill\pagebreak

\bibliographystyle{IEEEbib}
\bibliography{strings,refs}

\end{document}